# Towards Provable Secure Neighbor Discovery in Wireless Networks

Marcin Poturalski, Panos Papadimitratos, Jean-Pierre Hubaux
Laboratory for Computer Communications and Applications
EPFL, Switzerland
{marcin.poturalski, panos.papadimitratos, jean-pierre.hubaux}@epfl.ch

## ABSTRACT

In wireless systems, *neighbor discovery* (ND) is a fundamental building block: determining which devices are within direct radio communication is an enabler for networking protocols and a wide range of applications. To thwart abuse of ND and the resultant compromise of the dependent functionality of wireless systems, numerous works proposed solutions to *secure* ND. Nonetheless, until very recently, there has been no formal analysis of secure ND protocols. We close this gap in [24], but we concentrate primarily on the derivation of an impossibility result for a class of protocols. In this paper, we focus on reasoning about specific protocols. First, we contribute a number of extensions and refinements on the framework of [24]. As we are particularly concerned with the practicality of provably secure ND protocols, we investigate availability and redefine accordingly the ND specification, and also consider composability of ND with other protocols. Then, we propose and analyze two secure ND protocols: We revisit one of the protocols analyzed in [24], and introduce and prove correct a more elaborate challenge-response protocol.

## 1. INTRODUCTION

Wireless communication systems have been developed and deployed in increasing numbers and for diverse technologies, enabling a broad spectrum of applications. Nonetheless, although emerging wireless networking and mobile computing technologies offer a new rich set of tools, they also open the door to new vulnerabilities, primarily because wireless communication makes eavesdropping and injection of messages easy. Realizing that attacks against a wireless system can be perpetrated essentially anywhere and anytime, the research community devised a large volume of solutions to secure wireless networking protocols and applications.

Among these efforts, the problem of *securing neighbor discovery* has received significant attention. *Neighbor discovery* (ND), that is, discovering that a wireless device (node) is directly reachable without the assistance of any other device, is a fundamental element for practically any wireless system. As a result, an attack against ND, that is, misleading nodes into believing they are neighbors when they are not, does not merely introduce an artificial wireless communication link between the victim (misled) nodes. It is a simple method to compromise and abuse the system functionality that builds on ND: It is sufficient for the adversary to simply *relay* messages from one point to another in the network, and vice versa, without any message modification, to stage what is often termed a *wormhole* attack.

The consequences of such attacks can vary and be devastating. Consider, for example, the neighbor discovery of an access point (AP) performed by a mobile host in a WiFi network that allows access to networked resources. A relay attack may appear at first benign, or even helpful, as it essentially extends the AP range. However, it offers the adversary the opportunity to intercept and fully control communications of mobile hosts that cannot reach the AP but are 'attracted' to do so over the wormhole setup by the attacker. Similarly, for multi-hop wireless communication: In a wireless sensor network, nodes are likely to send their measurements to the sink over a path that includes a wormhole. In a different setting, a relay attack can be the most effective pick-pocket or lock-picking mechanism: The adversary approaches a victim user, carrying a radio frequency identification (RFID) tag for payments or for opening a car, garage, or office building door, she then relays communication between the RFID tag reader and the victim's tag, when the user is far from the RFID reader, and she gets a money charge through or physically accesses the building.

The basic approach against such relay attacks has been to protect the pair-wise execution of ND protocols with the help of various forms of *distance bounding* (DB) [3, 15, 8]: A node estimates its distance to another node by measuring the signal *time-of-flight (ToF)* to and/or from that node; if and only if the estimate is below a threshold, it declares it a neighbor. In spite of numerous schemes that are built on this approach, surveyed in Sec. 6, there has been no proof that secure ND is indeed achieved. This gap was brought forth very recently. [21] first argued informally that the existing proposals left the problem largely open and pointed out a common misconception in the notion of ND. Then, [24] established that, indeed, even for the most basic form of ND, what we term *two-party ND*, it is *impossible* for a large class of protocols that rely on time measurements to secure ND.

More important than the impossibility result per se, this highlights the need to prove the properties of security protocols. Developments over many years are not a substitute for rigorous reasoning on the protocol properties. In fact, a false feeling of trustworthy technology can lead to the deployment of solutions (algorithms, protocols) that were never proven secure. The subsequent discovery of attacks exploiting core design flaws rather than implementation glitches could only be a natural aftermath. The second important message that one can extract from [24] is the need to carefully take into consideration the idiosyncracies of the particular environment of the system to be secured. The basic observation behind the impossibility result is otherwise seemingly straightforward: Obstacles or interference can prevent nearby nodes from communicating directly, thus allowing the attacker to remain undetected while misleading two near-by nodes into believing they are neighbors (when they are not, precisely because of such obstructions).

An impossibility result, although it provides important insight into the problem at hand, does not provide solutions. Another crucial question is *proving the security of specific ND protocols* (under assumptions lifting the impossibility result, obviously). We make a first step in this direction in [24], proving correctness of a simple ND protocol. In this paper, we continue on this path. We refine and extend the framework of [24] (Sec. 2), providing a precise mathematical model of a wireless network and the adversary. These refinements and extensions bring our framework closer to the real world, in terms of reasoning on protocol correctness, and enable us to define the more elaborate *challenge-response (CR)* ND protocols we introduce in this paper (Sec. 3.2). We revisit one of the *beacon (B)* protocols considered in [24], and investigate additional protocols in our technical report [25]. Furthermore, we propose *more practical availability properties* (Sec. 3.1) than those geared towards and sufficient for the quest for the impossibility result. Clearly, a secure ND protocol concluding its execution correctly only 'once in a lifetime' satisfies the specification in [24] but it is not practical. In terms of further approaching practical instantiations, we also consider composability: We aim at results which hold when the secure ND protocols are used along with other protocols. All these elements lead to a precise problem specification, thus enabling us to develop proofs for secure ND protocols (Sec. 4). Before we conclude, we provide a discussion on the framework and the analysis performed in this paper, identify open problems (Sec. 5), and survey related work (Sec. 6).

# 2. SYSTEM MODEL

We are interested in modeling a wireless network: Its basic entities, *nodes*, are processes running on computational platforms equipped with transceivers communicating over a wireless channel. We assume that nodes have synchronized clocks (although not all protocols we consider in this paper make use of this assumption) and are static (not mobile). Nodes either follow the implemented system functionality, in which case we denote them as *correct*, or they are under the control of an adversary, in which case we denote them as *adversarial* nodes. Adversarial nodes can behave in an arbitrary fashion, also acting as correct nodes or lying dormant for any period of time.

We model communication at the physical layer rather than at higher layers (data link, network, or application), in order to capture the inherent characteristics of ND in wireless networks. For simplicity, correct nodes are assumed to use a single wireless channel and omnidirectional antennas, but we do not require them to have equal transmission power and receiver sensitivity. On the contrary, adversarial nodes use directional antennas to communicate across the wireless channel used by correct nodes, but they can also communicate across a dedicated *adversary channel* imperceptible to correct nodes.

Our system model comprises: (i) a *setting* $\mathcal{S}$ that describes the type (correct or adversarial) of nodes, their location and the state of the wireless channel; (ii) a *protocol model* $\mathcal{P}$ that determines the behavior of correct nodes; (iii) an *adversary model* $\mathcal{A}$ that determines the capabilities of adversarial nodes.

We assume that looking at the system at any point in time reveals one or more phenomena. We are interested in those relevant to the wireless communication and the system at hand and thus to our analysis. We denote these phenomena, associated with nodes, as *events* (Def. 3). Then, we model the system evolution over time using the notion of *trace*, i.e., a set of events (Def. 4). More precisely, we use *feasible* traces, which satisfy constraints specified by $\mathcal{S}$ (correspondence between wireless sending and receiving of messages), $\mathcal{P}$ (correct nodes follow the protocol), and $\mathcal{A}$ (adversarial nodes behave according to their capabilities). The constraints are defined by logical formulas we call *rules*.

## 2.1 System Parameters

Our model includes a number of parameters, listed below, which are determined by the technologies used by correct and adversarial nodes.

- $\mathbf{v} \in \mathbb{R}_{>0}$, the *signal propagation speed*, defining how fast messages propagate across the wireless channel, determined by the communication technology,
- $\mathbf{v}_{\text{adv}} \geqslant \mathbf{v}$, the *information propagation speed* over the *adversary channel*; as $\mathbf{v}_{\text{adv}} \geqslant \mathbf{v}$ this is also the maximum speed at which information can propagate,
- $\Lambda \subset 2^{\mathbb{R}^3}$, the set of *antenna patterns* that adversarial nodes can utilize with their directional antennas,
- $\Delta_{\text{relay}} \in \mathbb{R}$, the *minimum relaying delay* introduced by a node when relaying a message; this delay is due to processing exclusively, it does not include propagation time or any other delay.

Further, $\mathbb{V}$ denotes the set of unique *node identifiers*, which for simplicity we will consider equivalent with the nodes themselves.[1]

[1]Although this implies that every node is assigned a single identifier, it does not prevent an adversarial node from using (in the messages in sends) any identifier.

## 2.2 Settings

A setting describes the type and location of nodes, and how the state of the wireless channel changes over time.

DEFINITION 1. *A* setting $\mathcal{S}$ *is a tuple* $\langle V, loc, type, link, nlos\rangle$, *where:*

- $V \subset \mathbb{V}$ *is a finite set of nodes. An ordered pair* $(A, B) \in V^2$ *is called a* link.
- $loc : V \to \mathbb{R}^3$ *is the node* location *function. As we assume nodes are not mobile, this function does not depend on time. We define* $dist : V^2 \to \mathbb{R}_{\geqslant 0}$ *as* $dist(A, B) = d(loc(A), loc(B))$, *where d is the Euclidean distance in* $\mathbb{R}^3$. *We require the loc function to be injective, so that no two nodes share the same location. Thus,* $dist(A, B) > 0$ *for* $A \neq B$.
- $type : V \to \{correct, adversarial\}$ *is the* type *function; it defines which nodes are* correct *and which are* adversarial. *This function does not depend on time, as we assume that the adversary does not corrupt new nodes during the system execution. We denote* $V_{\text{cor}} = type^{-1}(\{correct\})$ *and* $V_{\text{adv}} = type^{-1}(\{adversarial\})$.
- $link : V^2 \times \mathbb{R}_{\geqslant 0} \to \{up, down\}$ *is the* link state *function. Accordingly, we say that at a given time* $t \geqslant 0$, *a link* $(A, B) \in V^2$ *is* up *(denoted* $t :: A \to B$*) or* down *(denoted* $t :: A \not\to B$*). We use abbreviations* $t :: A \leftrightarrow B =_{\text{def}} t :: A \to B \;\wedge\; t :: B \to A$ *and* $t :: A \not\leftrightarrow B =_{\text{def}} t :: A \not\to B \;\wedge\; t :: B \not\to A$. *We extend the "*$t :: A \to B$*" notation from single time points to sets as follows:* $T :: A \to B =_{\text{def}} \forall t \in T.\; t :: A \to B$. *We establish the convention* $\mathbb{R}_{\geqslant 0} :: A \not\to A$.
- $nlos : V^2 \to \mathbb{R}_{\geqslant 0}$ *is the* non-line-of-sight delay (NLOS) *function. If two nodes A and B can communicate over a line of sight, then* $nlos(A, B) = 0$. *Otherwise,* $nlos(A, B)$ *specifies the additional distance that the signal has to propagate compared to line-of-sight propagation* $dist(A, B)$. *We assume this function is symmetric, because of reciprocity of wireless links.*

*We denote the set of all settings by* $\Sigma$.

The ability to communicate directly, without the intervention or 'assistance' of relays, is expressed in our model by a link being up, thus the following definition:

DEFINITION 2. *Node A* is a neighbor *of node B in setting* $\mathcal{S}$ *at time t, if* $t :: A \to B$. *If* $t :: A \leftrightarrow B$ *we will say that nodes A and B* are neighbors *at time t.*

For simplicity in presentation, we use "$t :: A \to B$" to denote the neighbor relation and the link relation.

## 2.3 Message Space

The denote the set of all messages as $\mathbb{M}$. Any of the following is a message:

- an identifier $A \in \mathbb{V}$,
- a timestamp $t \in \mathbb{R}_{\geqslant 0}$,
- a location $l \in \mathbb{R}^3$,
- a nonce $n \in \mathsf{Nonces}$.

Moreover, two messages $m_1, m_2$ can be *concatenated* to form a message $\langle m_1, m_2\rangle$. Furthermore, an *authenticator* $\mathsf{auth}_A(m)$, where $A \in \mathbb{V}$ and $m \in \mathbb{M}$, is also a message.[2] Hence, messages are essentially terms, with the subterm relation is denoted by $\sqsubseteq$.

Every message $m$ has a *duration* $|m| \in \mathbb{R}_{\geqslant 0}$, which determines the transmission delay (*not* including the propagation delay), reflecting the bit-rate of the underlying communication technology. We assume that message duration is preserved by concatenation, but not by an authenticator. For $m = \langle m_1, m_2, \ldots, m_k\rangle$, the duration is $|m| = |m_1| + |m_2| + \ldots + |m_k|$ and the *position* of $m_i$ in $m$ is $pos(m_i \sqsubseteq m) = |m_1| + \ldots + |m_{i-1}|$, with $pos(m_1 \sqsubseteq m) = 0$; in the case of multiple occurrences of $m' \sqsubseteq m$, $pos(m' \sqsubseteq m)$ gives the position of the first occurrence. When we use the duration function for any concatenated message, we omit the brackets: $|m_1, m_2, \ldots, m_k|$. Finally, we assume that the duration of identifiers, timestamps, locations, nonces and authenticators in $\mathbb{M}$ is upper-bounded by some constant.

## 2.4 Events and Traces

We use the notion of *trace* to model an execution of the system. A trace is composed of *events*. We model events related to the wireless communication and the ND protocols operation. Each event is primarily associated with (essentially, takes place at) a node we call the *active* node.

DEFINITION 3. *An* event *is one of the following terms:*

- $\mathsf{Receive}(A; t; m)$
- $\mathsf{Bcast}(A; t; m)$
- $\mathsf{Dcast}(A; t; \alpha; m)$
- $\mathsf{Fresh}(A; t; n)$
- $\mathsf{Neighbor}(A; t; B, C, t')$
- $\mathsf{NDstart}(A; t)$
- $\mathsf{NDstart}(A; t; B)$

*where* $A \in \mathbb{V}$ *is the* active node, $t \in \mathbb{R}_{\geqslant 0}$ *is the event* start time, *denoted by start(.), and* $m \in \mathbb{M}$, $n \in \mathsf{Nonces}$, $\alpha \in \Lambda$, $B, C \in \mathbb{V}$, $t' \in \mathbb{R}_{\geqslant 0}$.

*Assuming that* $m_1 \sqsubseteq m_2$, *we use* $\mathsf{Bcast}(A; t; m_1 \sqsubseteq m_2)$ *to denote the event* $\mathsf{Bcast}(A; t - pos(m_1 \sqsubseteq m_2); m_2)$; *likewise for* Dcast *and* Receive.

The first three events are related to communication on the physical layer. Receive represents message reception. Bcast represents sending a message with an omnidirectional antenna. Dcast represents sending a message with a directional antenna using a pattern $\alpha \in \Lambda$. The pattern $\alpha$ is a subset of $\mathbb{R}^3$ indicating which nodes receive the message, assuming the sending node $A$ is located at $(0, 0, 0)$. We use the notation $B \in \alpha(A)$, meaning that $loc(B) - loc(A) \in \alpha$. The set of allowable antenna patterns, $\Lambda$, depends on the antenna used by the adversarial nodes. We do not dwell on the details of the structure of $\Lambda$, except for one requirement: $\mathbb{R}^3 \in \Lambda$; this is to ensure that adversarial nodes can use their antenna in an omnidirectional fashion.

Fresh is used to declare that nonce $n$ is (freshly) generated by $A$ at time $t$ or, in other words, that it was not sent before $t$. The remaining three events are specific to neighbor discovery protocols. Neighbor can be thought of as an internal outcome of a ND protocol (possibly reported to some higher layer): Node $A$ declares that $B$ is a neighbor of $C$ at time $t'$. Having $t'$ a single point in time is for simplicity only, and we could easily generalize to arbitrary sets. With NDstart, node $A$ declares that an instance of a ND protocol has been initialized: either with a specific node $B$ or with all neighbors. Next, traces comprising the above events are defined.

[2] The auth term represents an asymmetric authenticator, such as a digital signature.

DEFINITION 4. *A trace $\theta$ is a set of events that satisfies what we call the* finite cut condition*: for any finite $t \geqslant 0$, the subset $\{e \in \theta \mid start(e) < t\}$ is finite.*

The finite cut condition ensures that, during a finite amount of time, only a finite number of events occurs; as settings comprise a finite number of nodes, this is natural to demand.

We denote the set of all traces by $\Theta$, and $\Theta_{\mathcal{S},\mathcal{P},\mathcal{A}}$ the set of traces feasible with respect to a setting $\mathcal{S}$, a protocol $\mathcal{P}$ and an adversary $\mathcal{A}$.

## 2.5 Setting-Feasible Traces

The feasibility of a trace $\theta$ with respect to a setting $\mathcal{S} = \langle V, loc, type, link, nlos\rangle$ ensures a causal and strict time relation between send and receive events; it is formally defined by rules S1 – S4 (Fig. 1). Rule S1 ensures that every message that is received was previously sent. Dually, rules S2 and S3 ensure that a message broadcasted or sent with a directional antenna is received by all nodes enabled to do so by the link relation and, in the latter case, the antenna pattern used. In other words, communication is causal (a receive is always preceded by a sent), and reliable *as long as the link is up.* Unreliability, expected and common in wireless communications, is modeled by the state of the link being *down.* Furthermore, these rules introduce a strict time relation between events, reflecting the propagation delay from A to B, across the channel, with speed $\mathbf{v}$: $(dist(A,B) + nlos(A,B))\mathbf{v}^{-1}$. Rule S4 is a technical one: It ensures that no communication events are performed by nodes not present in setting $\mathcal{S}$, and that Bcast and Dcast events are used exclusively by correct and adversarial nodes, respectively. Note that this is not a restriction of the adversary: $\mathsf{Bcast}(A; t; m)$ can be emulated (i.e., trigger exactly the same Receive events) by $\mathsf{Dcast}(A; t; \mathbb{R}^3; m)$.

## 2.6 Protocol-Feasible Traces

Intuitively, a trace is feasible with respect to protocol $\mathcal{P}$ if correct nodes behave according to $\mathcal{P}$. Therefore the rules that specify this type of feasibility are protocol-dependent and are defined in Sec. 3.2. However, there is one general rule that dictates the behavior of correct nodes with respect to nonces. Rule F1 (Fig. 1) guarantees that if a nonce $n$ is freshly generated at time $t$ (i) the node that generated $n$ will not broadcast it *before* $t$, (ii) any other correct node who broadcasts a message containing nonce $n$ must have receive it (possibly in a different message) at least $\Delta_{\text{relay}}$ before broadcasting; this time difference is measured with respect to the positions of the nonce in the respective messages.

## 2.7 Adversary-Feasible Traces

We consider a single adversary model denoted by $\mathcal{A}$. Intuitively, adversarial nodes are allowed to send arbitrary messages, except for messages which would violate properties of authenticators or freshness.

A trace $\theta$ is feasible with respect to $\mathcal{A}$ if rules A1 - A2 (Fig. 1) are satisfied. Rule A1 deals with authenticators: An adversarial node is allowed to send a message containing arbitrary authenticators, as long as they are generated by an adversarial node (itself or other). This implies that adversarial nodes can share cryptographic keys or any material used for authentication. Furthermore, rule A1 reflects that the adversary cannot forge authenticated messages: Any message sent by an adversarial node that contains an authenticator generated by a correct node must be a relayed one. In other words, some (possibly the same) adversarial node must have received a message containing this authenticator earlier, at least $\Delta_{\text{relay}}$ plus the propagation delay between the two nodes over the adversarial channel. This condition reflects the structure of the adversarial channel: Any two adversarial nodes can establish direct communication. Rule A2 is similar to A1, but it is responsible for freshness: An adversary sending a message with a nonce generated by a correct can only be relaying the message (nonce). In this sense rule A2 is an adversarial equivalent of rule F1.

# 3. ND SPECIFICATION AND PROTOCOLS

In this section, we propose four types of ND protocols and their corresponding specifications. Due to limited space, we investigate in detail only two types; more can be found in our technical report [25]. We distinguish between (i) beacon-based protocols (*B-protocols*), represented by $\mathcal{P}^{\mathsf{B/T}}$, which require the transmission of one message by one of the protocol participants and synchronized clock for both participating nodes, and (ii) challenge-response protocols (*CR-protocols*), represented by $\mathcal{P}^{\mathsf{CR/TL}}$, which require a transmission of messages by both participants but *no* synchronized clocks. Within and across these categories, we distinguish protocols, as in [24], according to their capability to perform time measurements (T-protocols) or time measurements and location awareness (TL-protocols).

## 3.1 ND Properties

We consider two classes of properties ND protocols should satisfy. The first class pertains to *correctness* and consists of a single property, ND1 (Fig. 2): If two correct nodes[3] are declared neighbors at some time, then they must indeed be neighbors at that time. More precisely, there are two cases: (i) Node $A$ can declare that $B$ is its neighbor (i.e., $A$ can receive messages from $B$) or (ii) $A$ can declare that it is a neighbor of $C$ (i.e., $C$ can receive messages from $A$). In the latter case, property ND1 requires link $(C, A)$ to be up at not exactly time $t'$, but rather $dist(A, C) + nlos(A, C))\mathbf{v}^{-1}$ (propagation delay) after $t'$. As our model mandates that the link state is determined at the receiving end (node), if $A$ declares that it is a neighbor of $C$ at time $t'$, a message sent by $A$ at $t$ would be indeed received by $C$. In other words, $A$ is not forced to estimate the propagation delay to make a correct neighbor statement.

The second class of properties pertains to *availability*: If two nodes are neighbors for a long enough, protocol-specific time $T_{\mathcal{P}}$, the protocol must declare them neighbors. In the case of T-protocols, an additional notion needs to be introduced to formulate satisfiable availability properties: *neighbor discovery (ND) range*, $\mathbf{R} \in \mathbb{R}_{>0}$. Typically, $\mathbf{R}$ is equal to the *nominal communication range* for a given wireless medium and transceiver technology, however, we use $\mathbf{R}$ more

[3]The requirement that $B$ and $C$ be correct is explained in the Sec. 5.

S1 $\forall A \in V, t \in \mathbb{R}_{\geqslant 0}, m \in \mathbb{M}.\ \mathsf{Receive}(A;t;m) \in \theta \implies \exists B \in V.\ [t, t+|m|] :: B \to A$
$\wedge\ (\mathsf{Bcast}(B; t - (dist(A,B) + nlos(A,B))\mathbf{v}^{-1}; m) \in \theta\ \vee\ (\exists \alpha \in \Lambda.\ A \in \alpha(B)$
$\wedge\ \mathsf{Dcast}(B; t - (dist(A,B) + nlos(A,B))\mathbf{v}^{-1}; \alpha; m) \in \theta))$

S2 $\forall A, B \in V, t \in \mathbb{R}_{\geqslant 0}, m \in \mathbb{M}.\ \mathsf{Bcast}(B; t - (dist(A,B) + nlos(A,B))\mathbf{v}^{-1}; m) \in \theta$
$\wedge\ [t, t+|m|] :: B \to A) \implies \mathsf{Receive}(A;t;m) \in \theta$

S3 $\forall A, B \in V, t \in \mathbb{R}_{\geqslant 0}, m \in \mathbb{M}, \alpha \in \Lambda.\ (\mathsf{Dcast}(B; t - (dist(A,B) + nlos(A,B))\mathbf{v}^{-1}; \alpha; m) \in \theta$
$\wedge\ A \in \alpha(B)\ \wedge\ [t, t+|m|] :: B \to A) \implies \mathsf{Receive}(A;t;m) \in \theta$

S4 $\forall A, B \in \mathbb{V}, t \in \mathbb{R}_{\geqslant 0}, m \in \mathbb{M}, \alpha \in \Lambda.\ (\mathsf{Receive}(A;t;m) \in \theta \implies A \in V)$
$\wedge\ (\mathsf{Bcast}(A;t;m) \in \theta \implies A \in V_{\text{cor}})\ \wedge\ (\mathsf{Dcast}(A;t;\alpha;m) \in \theta \implies A \in V_{\text{adv}})$

F1 $\forall A, B \in V_{\text{cor}}, t_1, t_2 \in \mathbb{R}_{\geqslant 0}, n \in \mathsf{Nonces}, m_1 \in \mathbb{M}.\ n \sqsubseteq m_1\ \wedge\ \mathsf{Fresh}(A; t_1; n) \in \theta\ \wedge\ \mathsf{Bcast}(B; t_2; n \sqsubseteq m_1) \in \theta \implies$
$(A = B\ \wedge\ t_2 \geqslant t_1)\ \vee\ (A \neq B\ \wedge\ \exists \delta \geqslant \Delta_{\text{relay}}, m_2 \in \mathbb{M}.\ n \sqsubseteq m_2\ \wedge\ \mathsf{Receive}(B; t_2 - \delta; n \sqsubseteq m_2) \in \theta)$

A1 $\forall A \in V_{\text{adv}}, B \in \mathbb{V}, t \in \mathbb{R}_{\geqslant 0}, m, m_0, m_1 \in \mathbb{M}, \alpha \in \Lambda.\ m = \mathsf{auth}_B(m_0) \sqsubseteq m_1\ \wedge\ \mathsf{Dcast}(A; t; \alpha; m \sqsubseteq m_1) \in \theta \implies$
$(B \in V_{\text{adv}})\ \vee\ (\exists C \in V_{\text{adv}}, \delta \geqslant \Delta_{\text{relay}} + dist(C,A)\mathbf{v}_{\text{adv}}^{-1}, m_2 \in \mathbb{M}.\ m \sqsubseteq m_2\ \wedge\ \mathsf{Receive}(C; t - \delta; m \sqsubseteq m_2) \in \theta)$

A2 $\forall A \in V_{\text{cor}}, B \in V_{\text{adv}}, t_1, t_2 \in \mathbb{R}_{\geqslant 0}, \alpha \in \Lambda, n \in \mathsf{Nonces}, m_1 \in \mathbb{M}.\ n \sqsubseteq m_1\ \wedge\ \mathsf{Fresh}(A; t_1; n) \in \theta$
$\wedge\ \mathsf{Dcast}(B; t_2; \alpha; n \sqsubseteq m_1) \in \theta \implies \exists C \in V_{\text{adv}}, \delta \geqslant \Delta_{\text{relay}} + dist(C,B)\mathbf{v}_{\text{adv}}^{-1}, m_2 \in \mathbb{M}.\ n \sqsubseteq m_2$
$\wedge\ \mathsf{Receive}(C; t_2 - \delta; n \sqsubseteq m_2) \in \theta$

**Figure 1: Setting-, adversary- and common protocol-feasibility rules**

ND1 $\forall \mathcal{S} \in \Sigma, \theta \in \Theta_{\mathcal{S},\mathcal{P},\mathcal{A}}.\ \forall A, B, C \in V_{\text{cor}}, t, t' \in \mathbb{R}_{\geqslant 0}.\ \mathsf{Neighbor}(A; t; B, C, t') \in \theta \implies$
$(C = A\ \wedge\ t' :: B \to A)\ \vee\ (B = A\ \wedge\ (t' + (dist(A,C) + nlos(A,C))\mathbf{v}^{-1}) :: A \to C)$

$\mathsf{ND2}^{\mathsf{B/T}}$ $\forall \mathcal{S} \in \Sigma, \theta \in \Theta_{\mathcal{S},\mathcal{P},\mathcal{A}}.\ \forall A, B \in \mathbb{V}_{\text{cor}}, t \in \mathbb{R}_{\geqslant 0}.\ \mathsf{NDstart}(A; t) \in \theta\ \wedge\ [t, t + T_{\mathcal{P}}] :: A \to B$
$\wedge\ dist(A,B) + nlos(A,B) \leqslant \mathbf{R} \implies \exists t' \in [t, \infty), t'' \in [t, t + T_{\mathcal{P}}].\ \mathsf{Neighbor}(B; t'; A, B, t'') \in \theta$

$\mathsf{ND2}^{\mathsf{CR/TL}}$ $\forall \mathcal{S} \in \Sigma, \theta \in \Theta_{\mathcal{S},\mathcal{P},\mathcal{A}}.\ \forall A, B \in \mathbb{V}_{\text{cor}}, t \in \mathbb{R}_{\geqslant 0}.\ \mathsf{NDstart}(A; t; B) \in \theta\ \wedge\ [t, t + T_{\mathcal{P}}] :: A \leftrightarrow B$
$\wedge\ nlos(A,B) = 0 \implies \exists t_1, t_2 \in [t, \infty), t', t'' \in [t, t + T_{\mathcal{P}}].\ \mathsf{Neighbor}(A; t_1; A, B, t') \in \theta$
$\wedge\ \mathsf{Neighbor}(A; t_2; B, A, t'') \in \theta$

**Figure 2: Selected ND properties.**

freely as the communication range[4] for which ND inferences are drawn. In other words, nodes at a communication range larger than $\mathbf{R}$ will not be required to declare each other neighbor.

Fig. 2 displays two example ND2 properties for two types of protocols we consider: B/T-protocols and CR/TL-protocols. These two properties differ in four aspects, one depending on whether the protocol is T or TL, whereas the other three aspects depending on the protocol is beacon or challenge-response. The first aspect is the NDstart event: For CR-protocols, a particular neighbor $B$ with which ND is started is specified, whereas no such specification is necessary for B-protocols. Second, it may be required that link $(A, B)$ be up in only one direction (B-protocols) or both directions (CR-protocols). Third, for T-protocols an upper-bound on propagation distance in enforced ($dist(A,B) + nlos(A,B) \leqslant \mathbf{R}$), whereas for TL-protocols line-of-sight propagation is required ($nlos(A,B) = 0$). Forth, different forms of neighbor declaration are possible. The node making the declaration might be the same as (CR-protocols) or different (B-protocols) from the one initiating the ND protocol. Moreover the declaration might be uni-directional (B-protocols) or bi-directional (CR-protocols). Based on these differences, we can derive two additional ND2 properties $\mathsf{ND2}^{\mathsf{B/TL}}$ and $\mathsf{ND2}^{\mathsf{CR/T}}$(detailed in [25]).

## 3.2 ND Protocols

Fundamentally, beyond authentication mechanisms, all the ND protocols we consider measure the signal time-of-flight (ToF) between two nodes: B-protocols, with tightly synchronized clocks, are able to estimate ToF by transmitting a single beacon message, whereas CR-protocols require two messages, a challenge and a response, for the same purpose. T-protocols accept neighbor relations as valid if the ToF distance is below a threshold, as in [15], whereas TL-protocols require this distance to be equal to the geographical distance calculated based on nodes locations, as proposed in [24].

To make the presentation more approachable, we present the protocols in the form of pseudo-code, based on which we present the rules we developed to define the protocols. The pseudo-code is divided into *blocks* starting with a *triggering* event (**on** clause). If the triggering event occurs, the body of the block is executed, i.e., other events take place.

We start with a simple B/T-protocol we denote $\mathcal{P}^{\mathsf{B/T}}$, which is essentially the *temporal packet leash* protocol proposed by Hu, Perrig and Johnson in [15].

```
1: on NDstart(A; t1)
2:     Bcast(A; t1; ⟨A, t1, auth_A(t1)⟩)
3: on Receive(B; t2; ⟨A, t1, auth_A(t1)⟩)
4:     if t2 − t1 ⩽ R v^−1
5:         Neighbor(B; t2 + |A, t1, auth_A(t1)|; A, B, t2)
```

Block 1-2 describes the behavior after the ND protocol is started at node $A$ (e.g., by a higher layer protocol); P1 and P2 (Fig. 3) are the two rules that correspond to this block. Block 3-5 describes the behavior of a node after it receives a beacon message, and it is modeled by rules P3 and P4. Rule P1 is straightforward: if ensures that if the triggering event of block 1-2, $\mathsf{NDstart}(A; t_1)$, occurs in the trace, the event in the body of the block also occur. In the same fashion, rule P3 is defined for block 3-5, with an additional condition coming from the **if** clause.

These two rules are already sufficient to prove the ND2 property, but in a way, they only define half of aspects of the the protocol functionality. Indeed, nothing prevents at this point a node running this protocol from making arbitrary neighbor declarations. Rule P4 addresses this, stating that if a node makes a neighbor declaration, this has to be done according to block 3-5, i.e., the node had to receive a "fresh enough" beacon message. Only one aspect remains: Correct nodes are still allowed to broadcast arbitrary messages, including bogus beacon messages. This is addressed by rule P2. To motivate the definition of P2, let us consider an alternative rule would still be coherent with the pseudocode: If a correct node broadcasts a message at time $t_1$, this message is $\langle A, t_1, \mathsf{auth}_A(t_1)\rangle$. We can prove that such a defined protocol satisfies the ND specification. However, this is a weak result, precisely because that rule states that correct nodes cannot send any other messages than beacons. If the ND protocol were used along with or by any other protocol, obviously using other forms of messages, the result would no longer apply. To circumvent this undesired composability restriction, rule P2 is defined as follows. It only requires that if a correct node broadcasts at $t_1$ a message $m$ of a *particular form*, i.e., containing $\mathsf{auth}_B(t)$ as a subterm, then $m = \langle A, t_1, \mathsf{auth}_A(t_1)\rangle$. Hence, rule P2 gives a much less restrictive condition on protocols that can be securely composed with $\mathcal{P}^{\mathsf{B/T}}$: basically, it mandates that any other protocol does not use authenticated timestamps of this form.[5] Rule P4, in terms of composability, implies that the node cannot run any other ND protocol (i.e., a protocol making neighbor declarations), but we do not see this as a real restriction.

[4]By "communication range" we understand the actual distance plus NLOS effects.

Next, we describe $\mathcal{P}^{\mathsf{CR/TL}}$, a CR/TL-protocol. This protocol has a practical design twist: As authentication of a message can be time-consuming process, in this protocol we remove it from the time-critical ToF estimation phase. Otherwise, if the response needs too much time to be calculated, the clock of the challenging node can drift beyond an acceptable accuracy level. A protocol parameter $\Delta \in \mathbb{R}_{\geqslant 0}$ determines exactly how long after the challenge reception a node replies.

```
01: on NDstart(A; t1; B)
02:     Fresh(A; t1 + |B|; n1)
03:     Bcast(A; t1; ⟨B, n1⟩)
04: on Receive(B; t; ⟨B, n1⟩)
05:     Fresh(B; t + Δ; n2)
06:     Bcast(B; t + Δ; ⟨n2⟩)
07:     let τ > Δ
08:     Bcast(B; t + τ; ⟨loc(B), auth_B(n1, n2, loc(B))⟩)
09: on Receive(A; t; ⟨l, auth_B(n1, n2, l)⟩)
10:     if occurred Fresh(A; t1 + |B|; n1)
11:     if occurred Bcast(A; t1; ⟨B, n1⟩)
12:     if occurred Receive(A; t2; ⟨n2⟩)
13:     if v(t2 − t1 − Δ) = 2d(loc(A), l)
14:         Neighbor(A; t + |l, auth_B(n1, n2, l)|; A, B, t1)
15:         Neighbor(A; t + |l, auth_B(n1, n2, l)|; B, A, t2)
```

Note that we assume that a node keeps track of all the events it observes, and it can always refer to this 'history,' as in 10-12. Note also that there is no explicit block responsible for receiving the $\langle n_2\rangle$ response sent by $B$ in 06, because in this case node $A$ does not take any action other than recording the event occurrence, for later reference in line 11.

Considering again that "triggering event implies block body events," rule P1 is defined for block 01-03, P2 for block 04-08, and P4 for block 09-15. We do not define rules that restrict the occurrence of Fresh events (in lines 02 and 05) or the form of broadcasted messages (in lines 03 and 06), so that there is no obstacle for composability. For line 08, rule P3 is defined: If a node broadcasts a message $m$ containing a authenticator of the form $\mathsf{auth}_B(n_1, n_2, l)$, then $m$ precisely the message defined in line 08, and all the other events from block 04-08 occur. Finally, rule P5 is defined based on block 09-15. There is only one rule, despite two Neighbor events in lines 14 and 15, because both events match the universally quantified Neighbor event in P5; The rule uses a disjunction, as there are (small) timing differences in the node behavior depending on which of these two event is considered.

# 4. PROOFS

In this section we prove that two of the protocols we presented in the previous section satisfy the ND1 and ND2 properties. Before we proceed, we present two simple lemmas which facilitate subsequent proofs. Lem. 1 deals with authenticators and is an extension of rule A1, whereas Lem. 2 deals with freshness, extending S2 and F1. The proofs of these lemmas are similar, hence we only present the former.

Lemma 1. *Rule L1 (Fig. 4) holds for every trace $\theta$ feasible with respect to the adversary model $\mathcal{A}$ and some setting $\mathcal{S}$.*

Proof. The 1st disjunct ($B \in V_{\mathrm{adv}}$) follows immediately from A1, so we assume that $B \in V_{\mathrm{cor}}$ and focus on the 2nd disjunct, which we prove by contradiction: Fix $m = \mathsf{auth}_B(m_0)$. Assume that $\mathsf{Dcast}(A; t; \alpha; m \sqsubseteq m_1) \in \theta$, but $\mathsf{Bcast}(C; t'; m \sqsubseteq m_2) \notin \theta$, for any correct $C$, $t' \leqslant t - \Delta_{\mathrm{relay}} - \mathit{dist}(C, A)\mathbf{v}_{\mathrm{adv}}^{-1}$ and $m_2$ st. $m \sqsubseteq m_2$ ($\star$).

We use the following reasoning: Apply A1 to obtain $\mathsf{Receive}(D; t - \delta; m \sqsubseteq m_3) \in \theta)$, where $D \in V_{\mathrm{adv}}$ and $\delta \geqslant \Delta_{\mathrm{relay}} + \mathit{dist}(D, A)\mathbf{v}_{\mathrm{adv}}^{-1}$. Next, apply S1 to get $\mathsf{Dcast}(E; t - \delta - (\mathit{dist}(E, D) - \mathit{nlos}(E, D))\mathbf{v}^{-1}; \alpha'; m \sqsubseteq m_3) \in \theta$; the Bcast disjunct of S1 is ruled out by S4, assumption ($\star$) and $\mathbf{v}_{\mathrm{adv}} \geqslant \mathbf{v}$, as $\mathit{dist}(D, A)\mathbf{v}_{\mathrm{adv}}^{-1} + (\mathit{dist}(E, D) + \mathit{nlos}(E, D))\mathbf{v}^{-1} \geqslant \mathit{dist}(E, A)\mathbf{v}_{\mathrm{adv}}^{-1}$. This reasoning can be repeated ad infinitum, leading to an infinite number of Dcast events in $\theta$ with start time below $t$. This is a contradiction with the finite cut condition (Def. 4). Hence, $\mathsf{Bcast}(C; t'; m \sqsubseteq m_2) \in \theta$ for some correct $C$, $t' \leqslant t - \Delta_{\mathrm{relay}} - \mathit{dist}(C, A)\mathbf{v}_{\mathrm{adv}}^{-1}$ and $m_2$ st. $m \sqsubseteq m_2$. □

Lemma 2. *Rule L2 (Fig. 4) holds for every trace $\theta$ feasible with respect to the adversary model $\mathcal{A}$, some setting $\mathcal{S}$ and rule F1.*

Theorem 1. *If $T_{\mathcal{P}^{\mathsf{B/T}}} = \sup\{|A, t, \mathsf{auth}_A(t)| \mid A \in \mathbb{V}, t \in \mathbb{R}_{\geqslant 0}\} + \mathbf{R}\mathbf{v}^{-1}$ and $\Delta_{\mathrm{relay}} \geqslant \mathbf{R}\mathbf{v}^{-1}$ protocol $\mathcal{P}^{\mathsf{B/T}}$ satisfies ND1 and ND2$^{\mathsf{B/T}}$.*

Proof. First, we prove ND1 (Fig. 2). Consider a setting $\mathcal{S}$ and a trace $\theta \in \Theta_{\mathcal{S}, \mathcal{P}^{\mathsf{B/T}}, \mathcal{A}}$ such that $\mathsf{Neighbor}(B; t; A, C, t_2) \in \theta$ for $A, B, C \in V_{\mathrm{cor}}$. As $B$ is correct, apply P4 to get $C = B$ and $\mathsf{Receive}(A; t_2; \langle A, t_{,1}\, \mathsf{auth}_A(t_1)\rangle) \in \theta$, where $t = t_2 + |A, t_1, \mathsf{auth}_A(t_1)|$ and $t_2 \leqslant t_1 + \mathbf{R}\mathbf{v}^{-1}$ ($\star$). We need to show that $t_2 :: A \to B$.

Apply S1 to get $[t_2, t_2 + |A, t_1, \mathsf{auth}_A(t_1)|] :: D \to B$ and either:

[5] If this would pose a problem, the protocol can be modified, by e.g., authenticating a timestamp concatenated with some constant in place of simple the timestamp.

$\mathcal{P}^{\mathsf{B/T}}$

- P1 $\forall A \in V_{\text{cor}}, t_1 \in \mathbb{R}_{\geqslant 0}.\ \mathsf{NDstart}(A; t_1) \in \theta \implies \mathsf{Bcast}(A; t_1; \langle A, t_1, \mathsf{auth}_A(t_1)\rangle) \in \theta$
- P2 $\forall A \in V_{\text{cor}}, B \in \mathbb{V}, t_1, t \in \mathbb{R}_{\geqslant 0}, m \in \mathbb{M}.\ \mathsf{auth}_B(t) \sqsubseteq m \ \wedge\ \mathsf{Bcast}(A; t_1; m) \in \theta \implies m = \langle A, t_1, \mathsf{auth}_A(t_1)\rangle$
- P3 $\forall B \in V_{\text{cor}}, A \in \mathbb{V}, t_1, t_2 \in \mathbb{R}_{\geqslant 0}.\ \mathsf{Receive}(B; t_2; \langle A, t_1, \mathsf{auth}_A(t_1)\rangle) \in \theta \ \wedge\ t_2 - t_1 \leqslant \mathbf{Rv}^{-1}$
  $\implies \mathsf{Neighbor}(A; t_2 + |A, t_1, \mathsf{auth}_A(t_1)|; A, B, t_2) \in \theta$
- P4 $\forall B \in V_{\text{cor}}, A, C \in \mathbb{V}, t_2, t \in \mathbb{R}_{\geqslant 0}.\ \mathsf{Neighbor}(B; t; A, C, t_2) \in \theta \implies C = B$
  $\wedge\ \exists t_1 \in \mathbb{R}_{\geqslant 0}.\ \mathsf{Receive}(B; t_2; \langle A, t_1, \mathsf{auth}_A(t_1)\rangle) \in \theta \ \wedge\ t_2 - t_1 \leqslant \mathbf{Rv}^{-1} \ \wedge\ t = t_2 + |A, t_1, \mathsf{auth}_A(t_1)|$

$\mathcal{P}^{\mathsf{CR/TL}}$

- P1 $\forall A \in V_{\text{cor}}, B \in \mathbb{V}, t_1 \in \mathbb{R}_{\geqslant 0}.\ \mathsf{NDstart}(A; t_1; B) \in \theta \implies \exists n_1 \in \mathsf{Nonces}.$
  $\mathsf{Fresh}(A; t_1 + |B|; n_1) \in \theta \ \wedge\ \mathsf{Bcast}(A; t_1; \langle B, n_1\rangle) \in \theta$
- P2 $\forall B \in V_{\text{cor}}, t \in \mathbb{R}_{\geqslant 0}, n_1 \in \mathsf{Nonces}.\ \mathsf{Receive}(B; t; \langle B, n_1\rangle) \in \theta \implies \exists n_2 \in \mathsf{Nonces}, \tau > \Delta.$
  $\mathsf{Fresh}(B; t + \Delta; n_2) \in \theta \ \wedge\ \mathsf{Bcast}(B; t + \Delta; \langle n_2\rangle) \in \theta \ \wedge\ \mathsf{Bcast}(B; t + \tau; \langle loc(B), \mathsf{auth}_B(n_1, n_2, loc(B))\rangle) \in \theta$
- P3 $\forall B \in V_{\text{cor}}, C \in \mathbb{V}, t \in \mathbb{R}_{\geqslant 0}, n_1, n_2 \in \mathsf{Nonces}, l \in \mathbb{R}^3, m \in \mathbb{M}.\ \mathsf{auth}_C(n_1, n_2, l) \sqsubseteq m \ \wedge\ \mathsf{Bcast}(B; t; m) \in \theta \implies$
  $\exists \tau > 0.\ m = \langle loc(B), \mathsf{auth}_B(n_1, n_2, loc(B))\rangle \ \wedge\ \mathsf{Receive}(B; t - \tau - \Delta; \langle B, n_1\rangle) \in \theta$
  $\wedge\ \mathsf{Fresh}(B; t - \tau; n_2) \in \theta \ \wedge\ \mathsf{Bcast}(B; t - \tau; \langle n_2\rangle) \in \theta$
- P4 $\forall A \in V_{\text{cor}}, B \in \mathbb{V}, n_1, n_2 \in \mathsf{Nonces}, t_1, t_2, t \in \mathbb{R}_{\geqslant 0}, l \in \mathbb{R}^3.$
  $\mathsf{Receive}(A; t; \langle l, \mathsf{auth}_B(n_1, n_2, l)\rangle) \in \theta \ \wedge\ \mathsf{Fresh}(A; t_1 + |B|; n_1) \in \theta \ \wedge\ \mathsf{Bcast}(A; t_1; \langle B, n_1\rangle) \in \theta$
  $\wedge\ \mathsf{Receive}(A; t_2; \langle n_2\rangle) \in \theta \ \wedge\ \mathbf{v}(t_2 - t_1 - \Delta) = 2d(loc(A), l) \implies$
  $\mathsf{Neighbor}(A; t + |l, \mathsf{auth}_B(n_1, n_2, l)|; A, B, t_1) \in \theta \ \wedge\ \mathsf{Neighbor}(A; t + |l, \mathsf{auth}_B(n_1, n_2, l)|; B, A, t_2) \in \theta$
- P5 $\forall A \in V_{\text{cor}}, B, C \in \mathbb{V}, t, t_0 \in \mathbb{R}_{\geqslant 0}.\ \mathsf{Neighbor}(A; t; B, C, t_0) \in \theta \implies$
  $(C = A \ \wedge\ \exists n_1, n_2 \in \mathsf{Nonces}, t_1 \in \mathbb{R}_{\geqslant 0}, l \in \mathbb{R}^3.\ \mathsf{Fresh}(A; t_1 + |B|; n_1) \in \theta \ \wedge\ \mathsf{Bcast}(A; t_1; \langle B, n_1\rangle) \in \theta$
  $\wedge\ \mathsf{Receive}(A; t_0; \langle n_2\rangle) \in \theta \ \wedge\ \mathsf{Receive}(A; t - |l, \mathsf{auth}_B(n_1, n_2, l)|; \langle l, \mathsf{auth}_B(n_1, n_2, l)\rangle) \in \theta$
  $\wedge\ \mathbf{v}(t_0 - t_1 - \Delta) = 2d(loc(A), l)) \ \vee$
  $(B = A \ \wedge\ \exists n_1, n_2 \in \mathsf{Nonces}, t_2 \in \mathbb{R}_{\geqslant 0}, l \in \mathbb{R}^3.\ \mathsf{Fresh}(A; t_0 + |C|; n_1) \in \theta \ \wedge\ \mathsf{Bcast}(A; t_0; \langle C, n_1\rangle) \in \theta$
  $\wedge\ \mathsf{Receive}(A; t_2; \langle n_2\rangle) \in \theta \ \wedge\ \mathsf{Receive}(A; t - |l, \mathsf{auth}_C(n_1, n_2, l)|; \langle l, \mathsf{auth}_C(n_1, n_2, l)\rangle) \in \theta$
  $\wedge\ \mathbf{v}(t_2 - t_0 - \Delta) = 2d(loc(A), l))$

**Figure 3: Rules defining selected ND protocols.**

- L1 $\forall A \in V_{\text{adv}}, B \in V, t \in \mathbb{R}_{\geqslant 0}, m, m_0, m_1 \in \mathbb{M}, \alpha \in \Lambda.\ m = \mathsf{auth}_B(m_0) \sqsubseteq m_1 \ \wedge\ \mathsf{Dcast}(A; t; \alpha; m \sqsubseteq m_1) \in \theta \implies$
  $(B \in V_{\text{adv}}) \ \vee\ (\exists C \in V_{\text{cor}}, \delta \geqslant \Delta_{\text{relay}} + dist(C, A)\mathbf{v}_{\text{adv}}^{-1}, m_2 \in \mathbb{M}.\ m \sqsubseteq m_2 \ \wedge\ \mathsf{Bcast}(C; t - \delta; m \sqsubseteq m_2) \in \theta)$
- L2 $\forall A \in V_{\text{cor}}, B \in V, t_1, t_2 \in \mathbb{R}_{\geqslant 0}, \alpha \in \Lambda, n \in \mathsf{Nonces}, m \in \mathbb{M}.\ A \neq B \ \wedge\ n \sqsubseteq m \ \wedge\ \mathsf{Fresh}(A; t_1; n) \in \theta$
  $\wedge\ (\mathsf{Bcast}(B; t_2; n \sqsubseteq m) \in \theta \ \vee\ \mathsf{Dcast}(B; t_2; \alpha; n \sqsubseteq m) \in \theta) \implies t_2 \geqslant t_1 + dist(A, B)\mathbf{v}_{\text{adv}}^{-1} + \Delta_{\text{relay}}$

**Figure 4: Rules for Lemmas.**

(a) $\mathsf{Bcast}(D; t_2 - \delta_1; \langle A, t_1, \mathsf{auth}_A(t_1)\rangle) \in \theta \ \vee$

(b) $\mathsf{Dcast}(D; t_2 - \delta_1; \alpha; \langle A, t_1, \mathsf{auth}_A(t_1)\rangle) \in \theta$.

where $\delta_1 = (dist(D, B) + nlos(D, B))\mathbf{v}^{-1}$. Consider case (a). From S4 we get $D \in V_{\text{cor}}$ and then from P2 $D = A$. Thus $t_2 :: A \to B$, as desired.

Consider (b). Let $\tau = pos(\mathsf{auth}_A(t_1) \sqsubseteq \langle A, t_1, \mathsf{auth}_A(t_1)\rangle)$. Apply L1, to obtain for some $\delta_2 > \Delta_{\text{relay}}$ and $m$ st. $\mathsf{auth}_A(t_1) \sqsubseteq m$ that $\mathsf{Bcast}(E; t_2 + \tau - \delta_1 - \delta_2; \mathsf{auth}_A(t_1) \sqsubseteq m) \in \theta$. Refer to P2 to get $E = A$, $m = \langle A, t_1, \mathsf{auth}_A(t_1)\rangle$ and $t_1 = t_2 - \delta_1 - \delta_2 < t_2 - \Delta_{\text{relay}} \leqslant t_2 - \mathbf{Rv}^{-1}$. From the latter derive $t_2 > t_1 + \mathbf{Rv}^{-1}$. This is a contradiction with $(\star)$, thus (b) cannot be true. Consequently, (a) is the only valid option, and ND1 is satisfied.

Second, we prove $\mathsf{ND2}^{\mathsf{B/T}}$(Fig. 2). Consider a setting $\mathcal{S}$, where nodes $A, B \in V_{\text{cor}}$, $dist(A, B) + nlos(A, B) \leqslant \mathbf{R}$ and $[t_1, t_1 + T_{\mathcal{P}^{\mathsf{B/T}}}] :: A \leftrightarrow B$. Next, take any trace $\theta \in \Theta_{\mathcal{S}, \mathcal{P}^{\mathsf{B/T}}, \mathcal{A}}$ such that $\mathsf{NDstart}(A; t_1) \in \theta$. We need to show that $\mathsf{Neighbor}(B; t'; A, B, t'')$ for some $t' \geqslant t_1$ and $t'' \in [t_1, t_1 + T_{\mathcal{P}^{\mathsf{B/T}}}]$.

Start by applying P1 to $\theta$ to obtain $\mathsf{Bcast}(A; t_1; \langle A, t_1, \mathsf{auth}_A(t_1)\rangle) \in \theta$. As link $(A, B)$ is up for a sufficiently long time, S2 implies $\mathsf{Receive}(B; t_2; \langle A, t_1, \mathsf{auth}_A(t_1)\rangle)$, where $t_2 = t_1 + (dist(A, B) + nlos(A, B))\mathbf{v}^{-1}$. As $t_2 - t_1 \leqslant \mathbf{Rv}^{-1}$, P3 implies $\mathsf{Neighbor}(B; t_2 + |A, t_1, \mathsf{auth}_A(t_1)|; A, B, t_2)$. Obviously, $t' = t_2 + |A, t_1, \mathsf{auth}_A(t_1)| \geqslant t_1$ and $t'' = t_2 \in [t_1, t_1 + T_{\mathcal{P}^{\mathsf{B/T}}}]$, which completes the proof. □

Theorem 2. *If $\Delta_{\text{relay}} > 0$, $\mathbf{v}_{\text{adv}} = \mathbf{v}$ and $T_{\mathcal{P}^{\mathsf{CR/TL}}} = \infty$, protocol $\mathcal{P}^{\mathsf{CR/TL}}$ satisfies ND1 and $\mathsf{ND2}^{\mathsf{CR/TL}}$.*[6]

Proof. First, we prove ND1 (Fig. 2). Consider a setting $\mathcal{S}$ and a trace $\theta \in \Theta_{\mathcal{S}, \mathcal{P}^{\mathsf{CR/TL}}, \mathcal{A}}$ such that $\mathsf{Neighbor}(A; t; B, C, t_0) \in \theta$, where $A, B, C \in V_{\text{cor}}$. Apply P5 and arrive at two cases:

⟨I⟩ $C = A$: according to ND1, we need to prove $t_0 :: B \to A$

⟨II⟩ $B = A$: according to ND1, we need to prove $(t_0 + (dist(A, C) + nlos(A, C))\mathbf{v}^{-1}) :: A \to C$

We will consider both simultaneously. In case ⟨I⟩ P5 gives:

(1) $\mathsf{Bcast}(A; t_1; \langle B, n_1\rangle) \in \theta \ \wedge$

(2) $\mathsf{Fresh}(A; t_1 + |B|; n_1) \in \theta \ \wedge$

(3) $\mathsf{Receive}(A; t_2; \langle n_2\rangle) \in \theta \ \wedge$

(4) $\mathsf{Receive}(A; t_3; \langle l, \mathsf{auth}_B(n_1, n_2, l)\rangle) \in \theta \ \wedge$

(5) $\mathbf{v}(t_2 - t_1 - \Delta) = 2d(loc(A), l)$

for some $n_1, n_2 \in \mathsf{Nonces}, t_1, t_3 \in \mathbb{R}_{\geqslant 0}, l \in \mathbb{R}^3$ and $t_2 = t_0$ $(\star)$. In case ⟨II⟩, if we rename $C$ to $B$, P5 gives (1) – (5) for some $n_1, n_2 \in \mathsf{Nonces}, t_2, t_3 \in \mathbb{R}_{\geqslant 0}, l \in \mathbb{R}^3$ and $t_1 = t_0$ $(\star\star)$.

We continue by applying S1 to (4) to obtain for some $D \in V$:

[6] We set $T_{\mathcal{P}^{\mathsf{CR/TL}}} = \infty$ for simplicity: Otherwise, we would need to assume a maximum distance between $A$ and $B$ to have an upper-bound on the protocol execution time.

(a) $\mathsf{Bcast}(D; .; \langle l, \mathsf{auth}_B(n_1, n_2, l)\rangle) \in \theta \vee$

(b) $\mathsf{Dcast}(D; .; .; \langle l, \mathsf{auth}_B(n_1, n_2, l)\rangle) \in \theta$

("." in place of start time means that we are not concerned with the value.) Assuming (b), S4 implies $D \in V_{\text{adv}}$. Apply L1 to obtain $\mathsf{Bcast}(E; .; m)$ for some $E \in V$ and $m$ st. $\mathsf{auth}_B(n_1, n_2, l) \sqsubseteq m$ Then P3 gives for some $t_4 \in \mathbb{R}_{\geqslant 0}$:

(6) $\mathsf{Bcast}(B; .; \langle l, \mathsf{auth}_B(n_1, n_2, l)\rangle) \in \theta \wedge$

(7) $l = loc(B) \wedge$

(8) $\mathsf{Receive}(B; t_4 - \Delta; \langle B, n_1\rangle) \in \theta \wedge$

(9) $\mathsf{Bcast}(B; t_4; \langle n_2\rangle) \in \theta \wedge$

(10) $\mathsf{Fresh}(B; t_4; n_2) \in \theta$

The same is obtained under (a) via S4 and P3.
Apply S1 to (3) to get for some $F \in V$:

(11) $[t_2, t_2 + |n_2|] :: F \rightarrow A \;\wedge$
$(\mathsf{Bcast}(F; t''; \langle n_2\rangle) \in \theta \;\vee\; \mathsf{Dcast}(F; t''; .; \langle n_2\rangle) \in \theta)$

where $t'' = t_2 - (dist(F, A) + nlos(F, A))\mathbf{v}^{-1}$. We have two cases: (c) $F = B$ and (d) $F \neq B$. For case (c), given (10), F1 implies:

(c) $F = B \;\wedge\; t_4 \leqslant t_2 - (dist(A, B) + nlos(A, B))\mathbf{v}^{-1}$

In case (d), under (10), L2 implies $t_4 + dist(F, A)\mathbf{v}_{\text{adv}}^{-1} + \Delta_{\text{relay}} \leqslant t'' \leqslant t_2 - dist(F, A)\mathbf{v}^{-1}$. Using $\mathbf{v} = \mathbf{v}_{\text{adv}}$ and the triangle inequality we derive:

(d) $F \neq B \;\wedge\; t_4 \leqslant t_2 - dist(A, B)\mathbf{v}^{-1} - \Delta_{\text{relay}}$

Apply S1 to (8) to get for some $G \in V$:

(12) $[t_4 - \Delta, t_4 - \Delta + |\langle B, n_1\rangle|] :: G \rightarrow B \;\wedge$
$(\mathsf{Bcast}(G; t'''; \langle B, n_1\rangle) \in \theta \vee \mathsf{Dcast}(G; t'''; .; \langle B, n_1\rangle) \in \theta)$

where $t''' = t_4 - \Delta - (dist(G, B) + nlos(G, B))\mathbf{v}^{-1}$. Again, there are two cases: (e) $G = A$ and (f) $G \neq A$. In case (e), given (2), F1 implies:

(e) $G = A \;\wedge\; t_4 \geqslant t_1 + (dist(A, B) + nlos(A, B))\mathbf{v}^{-1} + \Delta$

In case (f), given (2), L2 implies $t_1 + |B| + dist(A, G)\mathbf{v}_{\text{adv}}^{-1} + \Delta_{\text{relay}} \leqslant t''' + |B| = t_4 - \Delta - (dist(G, B) + nlos(G, B))\mathbf{v}^{-1} + |B|$. After simple transformations using the triangle inequality and $\mathbf{v} = \mathbf{v}_{\text{adv}}$, and omitting the non-negative *nlos*:

(f) $G \neq A \;\wedge\; t_4 \geqslant t_1 + dist(A, B)\mathbf{v}^{-1} + \Delta + \Delta_{\text{relay}}$

We now have four possible cases to consider: (c)+(e), (c)+(f), (d)+(e) and (d)+(f). First, combine (5) and (7) to obtain:

(13) $t_2 - t_1 - \Delta = 2dist(A, B)\mathbf{v}^{-1}$

In case (c)+(e), after some simple transformations we obtain $t_2 - t_1 - \Delta \geqslant 2(dist(A, B) + nlos(A, B))\mathbf{v}^{-1}$. Given (13), this case is feasible if $nlos(A, B) = 0$. As $F = B$, (11) implies $t_2 :: B \rightarrow A$, which is what we needed to prove in case $\langle \text{I} \rangle$ given ($\star$). Furthermore, $G = A$ and (12) implies $(t_4 - \Delta :: A \rightarrow B)$. In case (e) $t_4 - \Delta = t_1 + (dist(A, B) + nlos(A, B))\mathbf{v}^{-1}$, which given ($\star\star$) means that property ND1 is also satisfied in case $\langle \text{II} \rangle$. Finally, as $\Delta_{\text{relay}} > 0$, it is easy to see that the remaining three cases are in contradiction with (13), which concludes the proof of ND1.

We now prove that $\mathsf{ND2}^{\mathsf{CR/TL}}$ holds. Assume that $\mathsf{NDstart}(A; t_1; B) \in \theta$. We need to prove $\mathsf{Neighbor}(A; t'_1; A, B, t') \in \theta \;\wedge\; \mathsf{Neighbor}(A; t'_2; B, A, t'') \in \theta$ for some $t'_1, t'_2 \in [t_1, \infty), t', t'' \in [t_1, t_1 + T_{\mathcal{P}^{\mathsf{CR/TL}}}]$.

First apply P1 to obtain $\mathsf{Bcast}(A; t_1; \langle B, n_1\rangle) \in \theta$. Next, as the link in up, $\mathsf{Receive}(B; t_2; \langle B, n_1\rangle) \in \theta$ is implied by S2, with $t_2 = t_1 + dist(A, B)\mathbf{v}^{-1}$ (the theorem assumes $nlos(A, B) = 0$). Apply P2 to get $\mathsf{Bcast}(B; t_2 + \Delta; \langle n_2\rangle) \in \theta$ and $\mathsf{Bcast}(B; t_2 + \tau; \langle loc(B), \mathsf{auth}_B(n_1, n_2, loc(B))\rangle) \in \theta$, where $\tau > 0$. Use the 'link is up' assumption and S2 to get $\mathsf{Receive}(A; t_4; \langle n_2\rangle) \in \theta$ and
$\mathsf{Receive}(A; t_5; \langle loc(B), \mathsf{auth}_B(n_1, n_2, loc(B))\rangle) \in \theta$, where $t_4 = t_2 + \Delta + dist(A, B)\mathbf{v}^{-1} = t_1 + \Delta + 2dist(A, B)\mathbf{v}^{-1}$. As $\mathbf{v}(t_4 - t_1 - \Delta) = 2dist(A, B)$ we conclude the proof by P4. □

# 5. DISCUSSION

We introduced a number of abstractions in our framework, simplifying wireless communications, for the sake of modeling and reasoning on secure ND (Sec. 5.1). In Sec.5.2, we outline here differences between protocols in terms of requirements and satisfied properties, and sketch open problems in Sec. 5.3.

## 5.1 Abstractions and Simplifications

**Mobility and NLOS Delay.** We assume nodes are static and NLOS delay constant over time. Otherwise, propagation delay would vary during the transmission of a message. In some cases, mobility and NLOS delay changes are negligible for the ND protocol execution time scale. For example, during $100\mu s$, nodes moving at 100kmph traverse $2.7mm$, which is below the accuracy of RF ranging systems (in the order of centimeters [1]). However, in general, mobility can have security implications. To see why, consider the $\mathcal{P}^{\mathsf{CR/TL}}$ protocol. If nodes move during the protocol execution, it is important when they estimate their location. At the very least, $A$ should estimate its location once when it sends the challenge, and again when it receives the response; whereas the responding node $B$ should estimate its location when it sends the response. But even this might be insufficient under high mobility: If $A$ measures its location at the beginning of the message, while $B$ measures the ToF at the end of the message, there may be space for a stealthy relay attack. Introducing mobility and a dynamically changing NLOS delay in our model is an interesting direction of our future work.

**Medium Access Control and Jamming.** For simplicity, we do not introduce any MAC restrictions into the model. Hence, a node is able to simultaneously receive any finite number of messages, even though in reality it is limited )to one message, or more for CDMA-like technologies). We could introduce additional rules that model radio interference, e.g., set links *down* if two (or more, depending on the node transceiver capabilities) simultaneous transmissions take place. However, this would *not* affect any of our results. Notably, the availability properties require links to be *up*, but they are agnostic as to why links are *up* or *down*. Similarly, jamming would not affect our results either: we capture jamming with links being *down*, thus availability implies, among other things, no jamming.

**Inaccuracies.** We assume correct nodes have accurate time and location information. However, in reality, inaccuracies are possible. Regarding time, clocks may be coarse-grained, they can drift, especially if the synchronization protocol fails, or the node may encounter difficulties in estimating message reception times over a noisy channel. Regarding location, infrastructure (e.g., Global Positioning System (GPS), or base stations) providing location information may be temporarily unavailable, or localization algorithms may be coarse grained. Some of the inaccuracies can be decreased: For example, averaging ToF over many messages decreases estimation errors. But some inaccuracy in time and location is unavoidable.

As secure ND protocols rely on distance estimates, their effectiveness can be affected by such inaccuracies. For T-protocols, and even more so for TL-protocols, inaccuracies hinder availability: they can lead to ToF estimates seemingly above the threshold for T-protocols, and make the two distance estimates diverge for TL-protocols. The only way to cope with these is to introduce some tolerance margins for measurements. Nonetheless, this would affect correctness: The higher the tolerance margin, the more space is left for fast relay attacks. This manifests the unsurprising tension between correctness and availability. Introducing inaccuracies explicitly into the framework is an interesting component for future work.

**Physical Layer Attacks.** The messages considered in our framework, albeit at the physical layer, are composed of "atomic" components, such as nonces and identifiers, typically assumed in formal security frameworks. In [7], Clulow et al. pointed out a number of physical layer attacks against DB protocols, working at the symbol (or bit) level. In the case of *external* adversaries, as considered in our ND specification, the attacks proposed in [7] can result in a (perceivably) negative $\Delta_{\mathrm{relay}}$. This can still be expressed in our model, hence our framework (notably the "atomicity" assumptions) is not limited with respect to those attacks. However, this is not the case for *internal* adversaries, which we discuss in the *Open Problems* section below.

## 5.2 Protocol Comparison

**T-protocols versus TL-protocols.** TL-protocols are less restrictive than T-protocols in term of correctness: They do not need the notion of ND range, $\mathbf{R}$, needed by T-protocols, and they are secure as long as $\Delta_{\mathrm{relay}} > 0$ (always true, unless th above-discussed inaccuracies and attacks come into play), while T-protocols require the $\Delta_{\mathrm{relay}}$ above $\mathbf{R}\mathbf{v}^{-1}$. In contrast, TL-protocols suffer in terms of availability: (i) they require location-aware nodes with secure location information, a far from trivial requirement, and (ii) they do not work for links with non-zero NLOS delay. We make a small note here on the nature of NLOS communication: Although there may be an obstacle between two nodes, it can still be possible to calculate the LOS message arrival time.[7] This, however, requires special care when selecting/designing a wireless receiver. Another practical disadvantage of TL-protocols is their requirement that the signal propagation speed be $\mathbf{v} = \mathbf{v}_{\mathrm{adv}}$ (Note: it is reasonable to assume $\mathbf{v}_{\mathrm{adv}} = c$, the speed of light); this limits TL-protocols to RF and other electromagnetic wave communications, while T-protocols can be used for lower speed technologies such as ultrasound.

**B-protocols versus CR-protocols.** B-protocols are conceptually simpler and have less stringent requirements for availability, requiring that links be *up* for shorter periods than those needed by CR-protocols. In contrast, B-protocols require tightly synchronized clocks, thus being impractical for many applications. In terms of correct (secure) operation, CRT-protocols require $\Delta_{\mathrm{relay}}$, the minimum relaying delaye, to be twice as large as that required by BT-protocols (for the same $\mathbf{R}$).

## 5.3 Open Problems

We observe that it is impossible to prove the correctness of the original 1993 distance bounding (DB) protocol by Chaums and Brands [3] either in our framework or in the framework by Meadows et al. [17] that deals with DB protocols. First, the Brands-Chaum protocol uses commitments and an XOR operation; although the XOR operation is modeled in [17], it cannot be used as it is in the Brands-Chaum protocol. More important, the Brands-Chaum protocol includes a *rapid-bit-exchange* (RBE) phase, during which nodes exchange single, fresh bits. This poses a problem for the usual modeling of freshness, that is, a message being fresh if it did not previously occur in a trace. Obviously, fresh RBE *bits* will repeatedly occur in a trace, as more than two will be exchanged.

The situation becomes even more interesting if we consider internal adversaries, that is, the execution of an ND (or DB) protocol with an adversarial node. In general, an attack is always possible: An adversarial node can collude with another adversarial node that is a neighbor of (is closer to) the victim node, and have it execute the ND or DB protocol on its behalf [3, 17]. However, with additional assumptions, for example, that "there is only one adversary node" [3] or that "nodes are prohibited to share cryptographic keys" [4]), some security guarantees can be claimed.

Nonetheless, in the presence of internal adversaries, the physical layer attacks in Clulow et al. [7] are much more significant than in the presence of external adversaries. We believe these types of attacks should be represented in any framework to prove the security of ND or DB protocols against internal adversaries. This would require a shift from a model that considers messages to one that considers physical communication layer symbols. Interestingly, this resembles the requirement to properly model the DB RBE. But this should not be a big surprise: RBE was introduced specifically to deal with internal attackers. The way to develop these models remains an open question.

# 6. RELATED WORK

The prevalent wormhole prevention mechanism is based on *distance bounding* (DB), which was first proposed by Brands and Chaum in [3] to thwart a relay attack between two correct nodes, also termed a *mafia fraud*. Essentially, DB estimates the distance between two nodes, with the guarantee that it is not smaller from their real distance. Subsequent proposals contributed in aspects such as mutual authentication [30], efficiency [9], operation in noisy environments [18, 28], and resistance to execution of the pro-tocol with a colluding group of adversarial nodes [4, 27]. In the latter, the attack termed *terrorist fraud* is thwarted under the assumption that adversarial nodes do not expose their private cryptographic material; if not, one adversarial node can undetectably impersonate another and successfully stage a terrorist fraud. *Authenticated ranging*, proposed by Čapkun and Hubaux in [29], lifts the technically non-trivial requirement of rapid response (present in all the above protocols), at the expense of not being resilient to a *distance fraud*, when the protocol is executed with a *single, non-colluding* adversarial node [4]. Finally, two other ND protocols that rely only on time measurements are the temporal packet leashes [15] (recall the $\mathcal{P}^{\mathsf{B/T}}$ is essentially a temporal packet leash) and TrueLink [8] (neither resistent to the distance fraud). The authors of [15] also proposed geographical packet leashes, which rely on nodes being location-aware. This protocol is quite similar to the TL-protocol introduced in [24]. But we emphasize the difference: the latter protocol requires clock synchronization as tight as that for temporal packet leashes, making it essentially a combination of temporal and geographical leashes, thus achieving secure ND in an environment with obstacles.

[7]If the LOS communication path is blocked by an obstacle, the LOS component of the received signal will be attenuated, and some NLOS components might arrive at the receiver with higher power. However, if the earlier-arriving LOS component is not too weak, with enough care it can be possible to detect it, calculate message reception time and the resultant LOS propagation delay.

A number of other secure ND schemes is proposed in the literature. Most of them rely on other wireless nodes or infrastructure, which may be (sometimes) unavailable, e.g., in WLAN or RFID systems. The approach of Poovendran and Lazos [23] relies on trusted, location-aware nodes (*guards*) to bootstrap ND. Hu and Evans have proposed a ND scheme utilizing properties of directional antennas in [14]. In [16], Maheshwairi et al. propose to use $k$-hop connectivity information obtained with a non-secure ND mechanism, and to inspect it for *forbidden structures*. Buttyan et al. also propose to use statistic of the connectivity graph, leading to a centralized framework, in [5]. Another centralized approach by Wang et al. [31] uses approximate distance measurements to visualize the network and enable a human operator to detect a wormhole attack. Finally, Rasmussen and Čapkun propose to use RF fingerprinting for secure ND [26]. Although this is a promising approach, it needs more practical investigation, notably about the feasibility of RF fingerprint forging.

The relay attack has been investigated in some recent works. One example is [10], where Hancke demonstrates a relay attack using only off-the-shelf hardware components, with a delay of around $20\mu s$. In [27] Reid et al. discuss using more sophisticated microwave repeaters to achieve nanosecond level relaying delays. We also refer an interested reader to a more theoretical work on relay (and other) physical layer attacks on DB by Clulow et al. [7], with a practical follow-up in [11], implementing some of these attacks against two commercial radio receivers used in RFID and sensor networks.

Recently, there has been a rising interest in formalizing analysis of security protocols in wireless networks. We mention works focusing on the security of routing [19, 2, 20, 32], local area networking [13], or broadcast authentication [12]. Closer to our work, the problem of DB has been treated formally in [17] by Meadows et al. Their paper is concerned with distance estimation rather than ND, but more importantly, the approach is different. The authors of [17] build on top of existing formal approaches [6, 22] tailored for "classical" security protocols, and augments it with a notion of distance based on time-stamps. However, it is not clear how neighborhood would be defined in this framework, nor how to model a protocol that uses location information, such as $\mathcal{P}^{\mathsf{CR/TL}}$. Beyond this, an interesting characteristic of their approach is that there is no explicit notion of an adversary. On the contrary, our approach starts with an explicit model of a wireless environment, including node location, state of wireless links, and an explicit adversary, controlling a number of nodes in the network. A potential advantage of this, although not shown in this paper, is that attack scenarios can be expressed in our model, whereas in [17] a collusion attack is described in an informal manner.

## 7. CONCLUSIONS

In this paper, we investigate how to analyze and design provably secure ND protocols, building on top of the framework introduced in [24]. We contribute a number of extensions that enable us to model and reason about more elaborate ND protocols (CR-protocols) than those previously considered (B-protocols). Basically, our revised framework (i) models additional practical aspects of wireless communications, (ii) caters to the co-existence and interoperability of secure ND protocols with other wireless security protocols, and (iii) focuses more than our work in [24] on sought properties that are of practical relevance, in particular, pertaining to the ND protocol availability.

We see this work as a step towards provably secure neighbor discovery. We outline a number of possible extensions to our framework, and open problems in the Discussion section. Among those, the seemingly most interesting one is to reason on secure ND protocols in the presence of internal adversaries. The nature of protocols that could deal with this type of adversarial behavior, as well as some recently discovered attacks [7], mandate, in our opinion, a shift from message-oriented to models that explicitly consider symbols at the physical communication layer.

## 8. ACKNOWLEDGMENTS

The authors would like to thank the anonymous reviewers for their helpful comments. The work presented in this paper was supported (in part) by the National Competence Center in Research on Mobile Information and Communication Systems (NCCR-MICS), a center supported by the Swiss National Science Foundation under grant number 5005-67322.